\documentclass[]{spie}  
\usepackage{amsmath,amsfonts,amssymb}
\usepackage{graphicx}
\usepackage[colorlinks=true, allcolors=blue]{hyperref}
\usepackage{upgreek}
\usepackage{epstopdf}
\usepackage{epsfig}

\usepackage{acronym}
\usepackage[caption=false]{subfig}

\acrodef{ALSI}[ALSI]{Advanced Laser Writing for Stellar interferometry}
\acrodef{AO}[AO]{Adaptive Optics}
\acrodef{AIP}[AIP]{The Leibniz Institute for Astrophysics Potsdam}
\acrodef{AWG}[AWG]{Arrayed Waveguide Grating}
\acrodef{DBC}[DBC]{discrete beam combiner}
\acrodef{DFG}[DFG]{Deutsche Forschungsgemeinschaft}
\acrodef{ELT}[ELT]{Extremely Large Telescope}
\acrodef{FBG}[FBG]{Fiber Bragg Grating}
\acrodef{FWHM}[FWHM]{Full Width at Half Maximum}
\acrodef{GLS}[GLS]{Gallium--Lantanum--Sulfide}
\acrodef{IFS}[IFS]{Integral Field Spectroscopy}
\acrodef{ISYS}[ISYS]{Institute for System Dynmaics Stuttgart}
\acrodef{KOOL}[KOOL]{the Koenigstuhl Observatory Opto-mechatronics Laboratory}
\acrodef{LBT}[LBT]{Large Binocular Telescope}
\acrodef{LSW}[LSW]{Landessternwarte}
\acrodef{MCF}[MCF]{multi-core Fiber}
\acrodef{MFD}[MFD]{mode-field diameter}
\acrodef{MM}[MM]{multi-mode}
\acrodef{MMF}[MMF]{multi-mode Fiber}
\acrodef{MLA}[MLA]{micro-lens array}
\acrodef{MOS}[MOS]{Multi Object Spectroscopy}
\acrodef{MPIA}[MPIA]{Max Plank Institute for Astronomy}
\acrodef{NA}[NA]{Numerical Aperture}
\acrodef{NAIR}[NAIR]{Novel Astronomical Instrumentation based on photonic light Reformating}
\acrodef{NIR}[NIR]{near-infrared}
\acrodef{P2VM}[P2VM]{pixel to visibility matrix}
\acrodef{PD}[PD]{Photonic Dicer}
\acrodef{PIMMS}[PIMMS]{Photonic Integrated Multimode Micro-Spectrograph}
\acrodef{PL}[PL]{photonic Lantern}
\acrodef{PSF}[PSF]{point spread function}
\acrodef{SM}[SM]{single-mode}
\acrodef{SMF}[SMF]{single-mode Fiber}
\acrodef{ULI}[ULI]{ultrafast laser inscription}
\acrodef{V2PM}[V2PM]{visibility-to-pixel-matrix}
\acrodef{PaL}[PaL]{path length}

\title{NAIR: Novel Astronomical Instrumentation through photonic Reformatting}

\author[a]{Robert J. Harris}
\author[b]{Jan Tepper}
\author[c]{John J. Davenport}
\author[c]{Ettore Pedretti}
\author[c]{Dionne M. Haynes}
\author[a]{Philipp Hottinger}
\author[a]{Theodoros Anagnos}
\author[c]{Abani Shankar Nayak}
\author[c]{Yohana Herrero Alonso}
\author[c]{Pranab Jyoti Deka}
\author[c]{Stefano Minardi}
\author[a]{Andreas Quirrenbach}
\author[b]{Lucas Labadie}
\author[c]{Roger Haynes}

\affil[a]{Zentrum f\"ur Astronomie der Universit\"at Heidelberg, Landessternwarte K\"onigstuhl, K\"onigstuhl 12, 69117 Heidelberg}
\affil[b]{I. Physikalisches Institut, Universit\"at zu K\"oln, Z\"ulpicher Str. 77, 50937 K\"oln, Germany}
\affil[c]{Leibniz-Institut f\"ur Astrophysik Potsdam, An der Sternwarte 16, D-14482 Potsdam, Germany}

\authorinfo{Further author information: (Send correspondence to R.J.H)\\E-mail: rharris@lsw.uni-heidelberg.de\\ Website: https://www.lsw.uni-heidelberg.de/projects/NAIR/cmsms/}

\begin{document} 
\maketitle

\begin{abstract}
The project "\acl{NAIR}" is a DFG-funded collaboration to exploit the recognized potential of photonics solutions for a radically new approach to astronomical instrumentation for optical/infrared high precision spectroscopy and high angular resolution imaging. 
We present a project overview and initial development results from our Adaptive Optics-photonic test bed, Ultrafast Laser Inscribed waveguides for interferometric beam combination and 3D printing structures for astronomical instrumentation. 
The project is expected to lead to important technological breakthroughs facilitating uniquely functionality and technical solutions for the next generation of instrumentation.

\end{abstract}


\section{Introduction}
\label{sec:intro}  
\acresetall

Photonic reformatting can be defined as an instrumental technique that utilizes photonic components to improve performance through a spatial rearrangement or reformatting of the distribution of starlight collected by an astronomical telescope, either in the pupil or in the image plane. 

In its simpler form, a reformatter is a network of waveguides or fibers used to remap the spatial arrangement of input ports into an equivalent number of output ports. Examples are fiber networks used in \ac{MOS} \cite{Hill:1980} or \ac{IFS} \cite{Vanderriest:1980} to reformat the fibers sampling the focal plane of the telescope into a pseudo-slit feeding the spectrograph. \acp{PL} \cite{Leon-Saval:2005} are more advanced forms of reformatters, which are tapered multi-waveguide/fiber devices, that transform an $N$-modes \ac{MM} waveguide/fiber into $N$ output \ac{SM} waveguides/fibers with minimal light losses. 

Applications of photonic reformatters can be broadly classified according to the coherent or incoherent level of optical field detection required at the output of the device. In this context, incoherent reformatting implies that only the light intensity of the collected signal is measured, whereas coherent reformatting implies that both the amplitude and phase are the measurable quantities. 

Coherent reformatters are needed for applications requiring coherent field detection such as high angular resolution astronomy where phase information is crucial. Examples include long-baseline interferometry, sparse aperture interferometry \cite{Haniff:1987} and its improved derivative pupil-remapping interferometry \cite{Perrin:2006}. The paradigmatic examples in the latter case are pupil plane reformatters that use optical-path-equalized waveguides to rearrange light, collected by a 2D array of sub-pupils, into a non-redundantly spaced waveguide array, suitable for multi-axial, interferometric beam combination schemes \cite{Jovanovic:2012}. In this case, the critical component functionality is the capability to preserve the phase relationship between the light collected by the input sub-pupils.

Applications requiring incoherent optical field detection are most common in spectroscopy. To this category belong components like 1) mode scramblers \cite{Haynes:2014} and 2) photonic reformatters \cite{Harris:2015}, both based on the \ac{PL} concept \cite{Leon-Saval:2005}. In these cases, the function of the reformatter is to 1) reduce the modal noise in a \ac{MM} fiber-fed spectrograph through a rapid and efficient phase scrambling of the output modes of the output \ac{MM} fiber, 2) increase the intensity scrambling of characteristics, thereby reducing the generation of spurious spatial distributions ("print through"), and 3) reformat the input 2D \ac{MM} light distribution into a 1D pseudo-slit featuring a diffraction limited profile in the dispersion direction of the spectrograph.

In these cases, loss of the phase information contained in the input field of the device is crucial to achieve a uniform, time-averaged \ac{PSF} of the spectrograph, as required by high-resolution spectroscopic applications.

We break our project up into three of areas of work, these are listed below:

\textbf{Sparse aperture interferometry}: Pioneered at optical and infrared wavelengths by Ref. \citenum{Haniff:1987} and \citenum{Tuthill:2000}, aperture masking treats the primary mirror as multi-aperture non-redundant interferometer that measures the visibilities and the closure phases, from which an image can be reconstructed. Non-redundancy of the sub-array is essential to recover an unperturbed estimate of the phase of the object at given spatial-frequency. This technique has produced astrophysical images with exquisite fidelity and $\lambda$/(2$B$) resolution ($B$ being the maximum baseline of the sub-array), and represents a highly competitive method to retrieve diffraction-limited images with seeing- limited telescopes.
A well-known limitation of aperture masking is that it uses only between 1 \% and 10 \% of the photons, due to constraints imposed by the non-redundancy condition and the coherence length of the atmosphere. This biases the technique to bright sources. In order to avoid wasting a significant fraction of the light entering the telescope, \cite{Perrin:2006} proposed the so-called pupil remapping technique, a derivation of the aperture masking technique. It is based on the use of photonic devices (e.g. a \ac{SMF} bundle) to transform or remap the fully redundant input pupil of the telescope into a non-redundant output pupil used to measure interferometric visibilities and closure phases for high-fidelity, high-resolution image reconstruction. In practice, the difficulty of calculating a fully non-redundant pupil for more than 15 sub-apertures and the large number of required encoding pixels (with consequent high readout noise) has limited the efficiency of this approach to bright sources, partly because of the poor mechanical stability of fibers at visible wavelengths \cite{Huby:2012}.
With the recent progress made in interferometric integrated optics beam combining devices, new designs based on multi-combiner concepts as studied in Ref. \citenum{Minardi:2012c} can be implemented to extend sparse aperture interferometry to the full pupil.

\textbf{Photonic reformatters}: To achieve higher spectral resolving power the input fibers to spectrographs are frequently sub-divided by an image slicer. This allows an increase of spectral resolution, but requires another optical element, which reduces throughput and increases complexity. We propose to reformat the shape of the fiber within the fiber, which simplifies the alignment process and potentially increases throughput. Several different shapes for reformatters have been proposed and trialled (e.g. Ref. \citenum{Leon-Saval:2012, Harris:2014, Cvetojevic:2013}). We will concentrate on those that reformat the \ac{PSF} into a long slit in a similar way to conventional spectrographs. This allows for ease of retrofitting to conventional spectrographs if required. In addition to reformatting we will trial \ac{SM} (diffraction-limited) slits. These are diffraction limited in the dispersion direction, potentially removing the problem of modal noise in the spectra. However whether this is the case still needs to be proven, so the project will be a great addition to this.

\textbf{Scrambling Fibers}: Precision spectroscopy in astronomy, such as exoplanet detection by precision radial velocity measurement techniques, is a rapidly developing area of astronomical research, and there is a strong drive to detect habitable Earth size planets. There are several new precision spectroscopy facilities under development or recently completed, e.g. ESPRESSO \cite{Pepe:2010}, PEPSI \cite{Strassmeier:2015}, CARMENES \cite{Quirrenbach:2010, Quirrenbach:2014} and SPIRou \cite{Thibault:2012}. Some of these instruments operate in the \ac{NIR} band (e.g. ELT-HIRES, \cite{Zerbi:2014} for the \ac{ELT}, GIANO \cite{Origlia:2014} for TNG and HZPF \cite{Mahadevan:2014} for HET) and are designed to determine radial velocity shifts of the order of 1 m/s or less. This not only requires a very stable spectrograph but also (if fiber fed) a very stable fiber feed. However, \ac{MM} optical fibers suffer from modal noise and this has been shown to critically limit the signal-to-noise ratio achievable in fiber-coupled, high-resolution spectrographs. Because modal noise scales with increasing wavelength (i.e. decreasing mode numbers), it represents a limitation for the current and next generation of near-infrared precision radial velocity spectrographs. The scrambling fiber development that we are proposing, based on the \ac{MCF} and \ac{PL} technologies \cite{Haynes:2014}, aims at significantly improving scrambling and reducing modal noise in the optical fibers linking both the telescope and precision calibration system to the spectrograph, thus boosting the performance of future instruments.

The \ac{NAIR} consortium aims to develop reformatting in all the described areas. Our project was funded in January 2017 by the \ac{DFG} and in this paper we summarize the work performed so far. In Section \ref{sec:coherent} and \ref{sec:incoherent} we report on the initial  coherent and incoherent results. In Section \ref{sec:future} we discuss our future work and conclude in Section \ref{sec:conclusions}. 

\section{Work on coherent reformatting devices}
\label{sec:coherent}
The combination of multiple beams within an integrated optics chip has proven to be an efficient solution in the context of both long baseline interferometry \cite{GRAVITY} as well as aperture masking interferometry \cite{Jovanovic:2012}. Besides their small footprint, \ac{SM} waveguides naturally operate as a spatial filter leading to a well-defined instrumental transfer function. As a consequence, integrated optics couplers allow the measurement of interferometric visibilities with a much higher accuracy than conventional bulk optics solutions. However, the coherent on-chip combination poses sensitive requirements in terms of differential dispersion and polarization properties and needs to compete with the high throughput of classical bulk optics. Our research efforts in this field concentrate on innovative on-chip beam combination schemes such as reformatters and \acp{DBC} as well as exploring new materials to access the 3-5\,$\upmu$m range. 

\subsection{Integrated 4- Telescope coherent Reformatters} 
\label{subsec:4TR}
Coherent Reformatters are specially designed waveguides that rearrange between one set of positions and another whilst retaining the same path length. The tolerance required for the path length difference depends on the observing wavelength as the Fried parameter describing the atmospheric turbulence depends on $\sim$ $\lambda ^\frac{6}{5}$. The construction of these reformatters or special waveguides is motivated by Ref. \cite{charles} using spline technique. These spline waveguides had been successfully implemented in the second generation Dragonfly instrument to achieve experimental throughputs of 70 \% at the output of the waveguides \cite{Norris:2014}. 

\begin{figure}
  \centering
  \begin{tabular}{cc@{}}
  \subfloat[]{\includegraphics[width=0.45\textwidth]{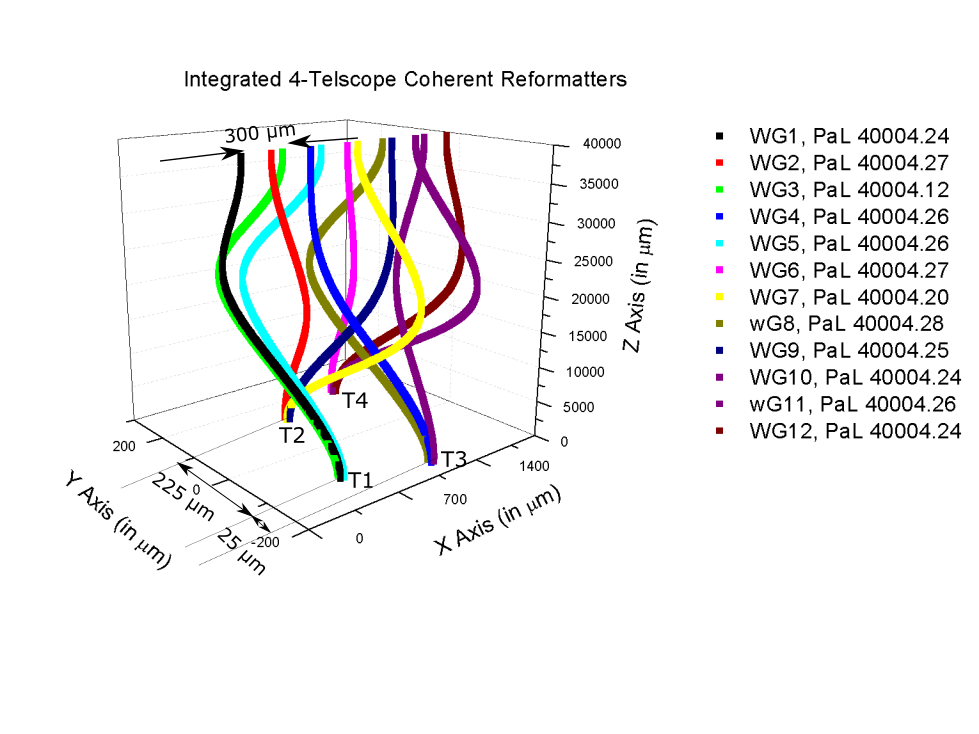}\label{4ta}}
  \subfloat[]{\includegraphics[width=0.5\textwidth]{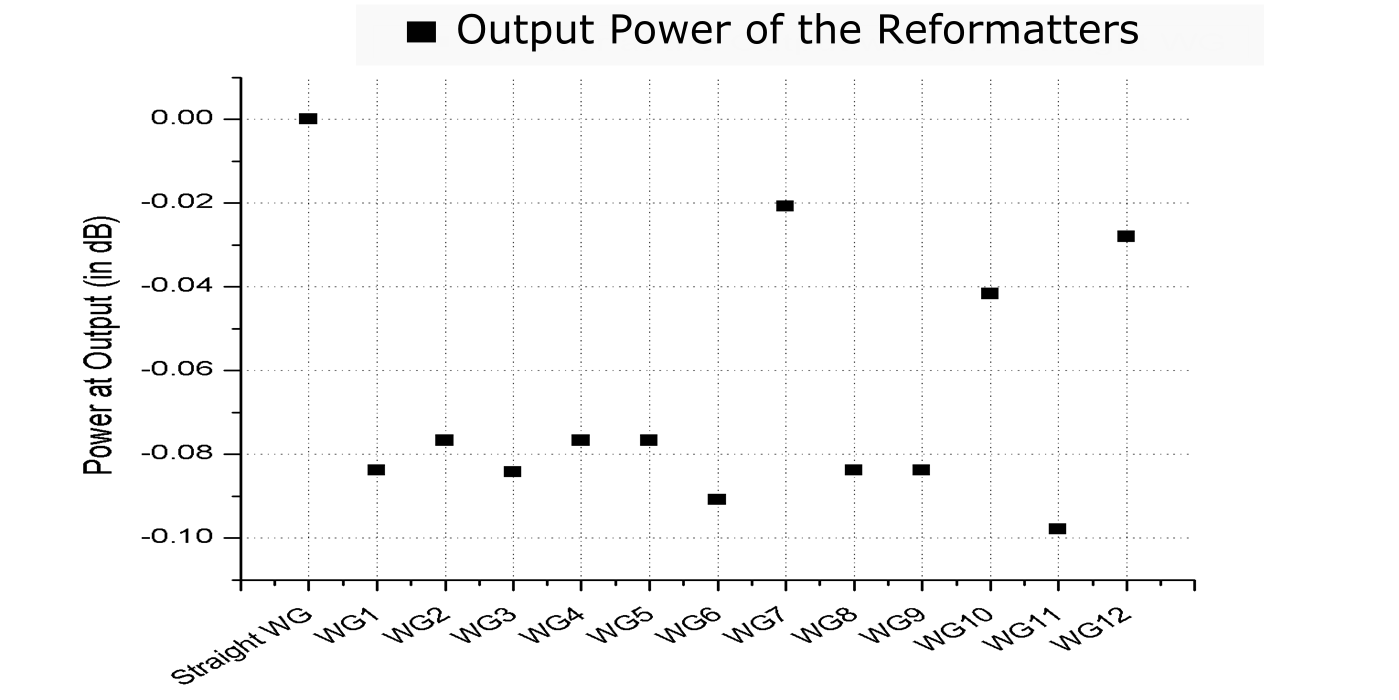}\label{4tb}}
  \end{tabular}
  
  \caption{\textbf{a)} 3D reformatting similar to planar geometry proposed by M. Benisty et. al. \cite{Benisty2009}.  There is an offset of 25 $\upmu$m at the input and a period of 300 $\upmu$m at the output. $\Delta$$L$ in between all the pair of waveguides is 0.1 $\upmu$m. Here PaL:Path Length, T:Telescope. \textbf{b)} Power at the output of all the 12 waveguides. All the powers are normalized w.r.t a straight waveguide.}
  \label{4t}
\end{figure}

Fig. \ref{4ta} shows the design of a 3D reformatter created using spline technique, it is similar to the idea of pairwise beam combination \cite{Benisty2009}. The planar design had unnecessary waveguide crossings that gave to undesirable cross-talk and we believe that it can be mitigated using 3D geometry. Light from 4 telescopes is injected at the input of a single chip and are divided by tri-couplers arranged in a triangular geometry. These are then combined interferometrically giving 6 pairs, a total of 12 outputs. The outputs from the pairwise waveguides can be combined in free-space or using integrated beam combination mechanism on the same chip to retrieve the complex visibilities. The input position for the 4 telescopes are offset by 25 $\upmu$m and the 6 pairwise outputs of the waveguides are kept at a period of 300 $\upmu$m to minimize the amount of stray light. The coherence length of the photons under turbulent atmosphere at 3.4 $\upmu$m is roughly $\approx$ 30\,$\upmu$m \cite{Jovanovic:2012}. Hence, for our case as seen from Fig. \ref{4ta}, it is sufficient to have path length difference, $\Delta$$L$ of 0.1 $\upmu$m  between all the pairs of waveguides. 

We evaluated the performance of these spline waveguides using BEAMPROP in RSOFT. We used a step index profile with $n_{\rm{core}}$ = 2.315, $n_{\rm{cladding}}$ = 2.31 and a channel waveguide geometry of 8.5 $\upmu$m$\times$ 17 $\upmu$m that is \ac{SM} at $\lambda$ of 3.4 $\upmu$m\cite{Diener2017,dienerspie}. We measured the output of all spline waveguides and normalized it w.r.t a straight waveguide. We can see from Fig. \ref{4tb} that the average power of all waveguides within a certain discrepancy is around -0.08 dB that leads to theoretical normalized throughput of $\approx$ 98\%. Our model does not include material dispersion and the minimum radius of curvature $(R_{\rm{c}})$ for all the waveguides was more than 1000 mm due to the device length of 40 mm. Our study also showed that the throughputs dropped by more than 50\% when $R_{c}$ was $\leq$ 16 mm or when the length of the device was 25 mm for the same configuration. Hence, for the proposed geometry there were minimal bend and transition losses and the throughput of these waveguides was much higher in simulations. This will not be the case when practical devices will be fabricated using \ac{ULI} \cite{Arriola:2013} and will need to be addressed in the future.         

\subsection{Discrete beam combiners}
\label{subsec:dbc}

We are investigating novel beam combination schemes that go beyond the classical 2x2 or ABCD couplers \cite{Benisty2009}.  In particular, we focus on so-called \acp{DBC} \cite{Diener2017} in which a large set of evanescently coupled waveguides produces an interferometric output pattern from which the mutual visibilities of the input beams can be derived. Mathematically, the relationship between the input visibilities and the output intensity pattern is described by the \ac{V2PM} \cite{Saviauk2013}. The quantities of interest, i.e. the mutual coherence functions of the input beams, can be obtained by measuring the output patterns and applying the inverse \ac{V2PM} (then called \ac{P2VM}). The challenge is to obtain a \ac{DBC} configuration for which this problem is well conditioned, i.e. insensitive to small errors in the intensity measurements.

A promising realization of a \ac{DBC} is the so-called zig-zag array, see Fig.\,\ref{fig:coherent_right}. Such a configuration was tested with monochromatic light at 1.30, 1.50 and 3.39 $\upmu$m. The advantage of this type of beam combiner is that it does not require any bends in the waveguides and therefore eliminates any bend losses. Naturally, any device based on evanescent coupling exhibits chromatic behavior. Therefore, also the \ac{V2PM} of the zig-zag array is expected to be chromatic. We are currently working on the first characterizations with broadband light to progress towards a first on-sky demonstration.

\subsection{6- Telescope discrete beam combiner for the H band}
\label{subsec:DBC-H}
We have manufactured 6-telescope discrete beam combiners in \ac{GLS} using \ac{ULI}. The combiners were manufactured to demonstrate the performance of components that could be hosted at current interferometric facilities, although for the \ac{NAIR} project we only use 4-input beam combiners. Since the \ac{ULI} manufacturing process requires several steps of calibration we have manufactured a test facility that permits the characterization of a beam combiner of 6-telescope complexity in a couple of minutes per wavelength. We then use a tunable laser to characterize the beam combiner across the whole bandwidth. 

\subsubsection{Automated  J and H band DBC characterization bench}
\label{subsec:J-H_testbed}

\begin{figure}[b]
 \begin{tabular}{ll@{}}
 \subfloat[]{\includegraphics[width=0.47\textwidth]{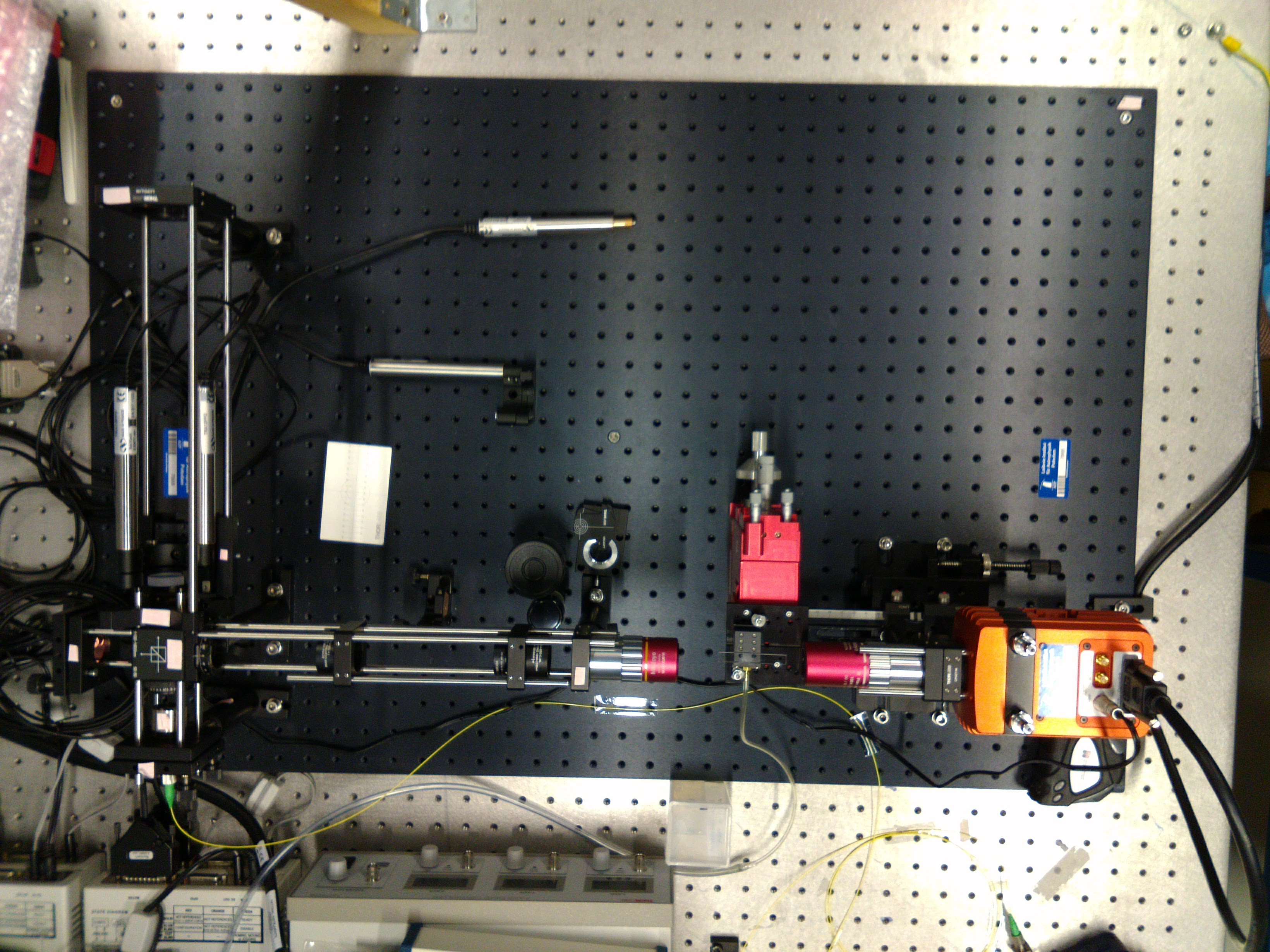}\label{JH-testbench}}
 & \subfloat[]{\includegraphics[width=0.47\textwidth]{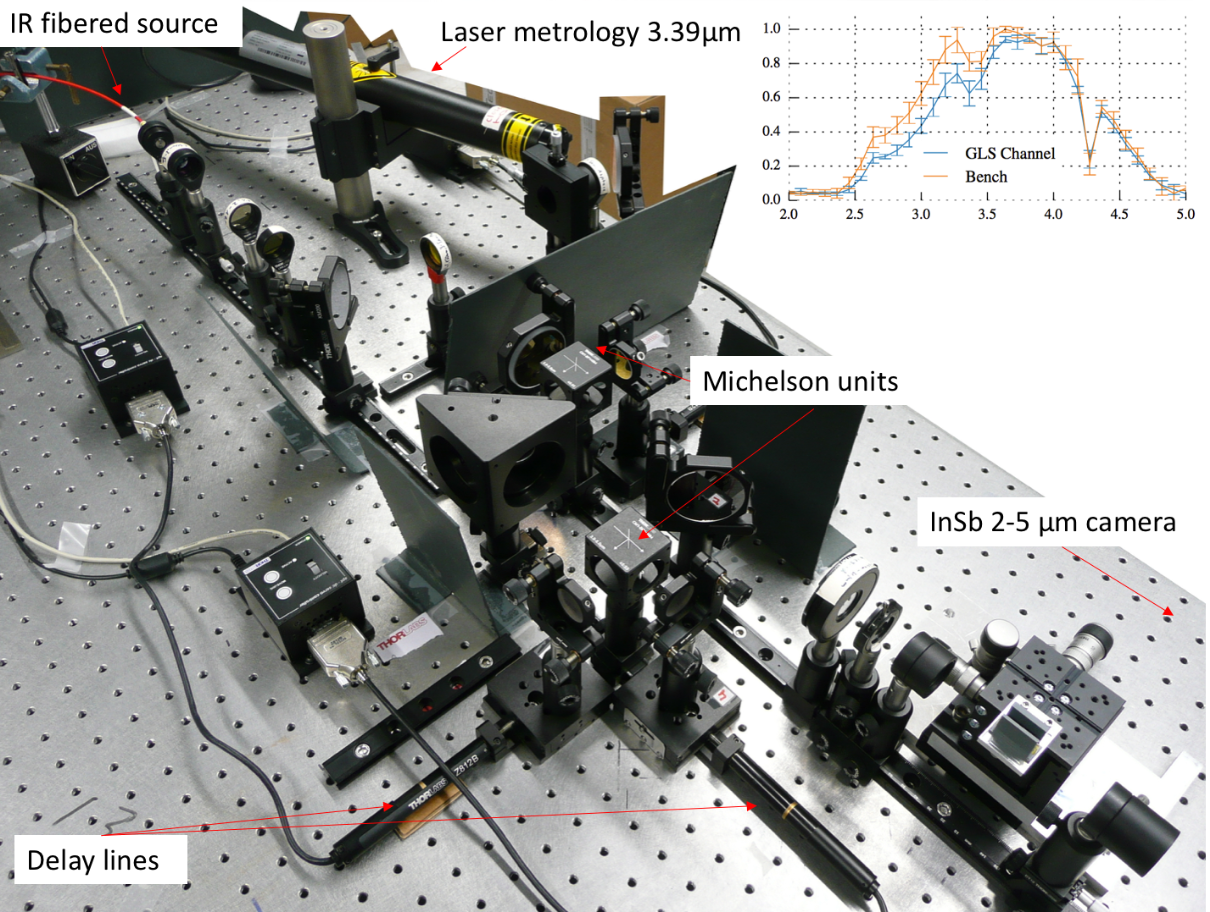}\label{Koeln-testbench}}
 \end{tabular}
\caption{\textbf{a)} The J-band and H-band DBC characterization bench. The picture shows the current two beams setup at innoFSPEC in Potsdam. \textbf{b)}  Mid-infrared 4-telescope characterization bench developed in Cologne and covering the 2-5 $\upmu$m range (the picture only shows the L and M band channel). A direct-spectroscopy capability is currently being added to the bench. The inset shows an example of a mid-infrared spectrum currently retrieved by FTS between 2 and 5 $\upmu$m where the CO$_{2}$ absorption band is visible at 4.2 $\upmu$m}.
\end{figure}

\begin{figure}[b]
 \begin{tabular}{ll@{}}
 \subfloat[]{\includegraphics[width=0.5\textwidth]{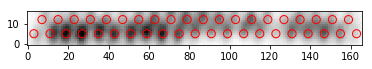}}
 & \subfloat[]{\includegraphics[width=0.5\textwidth]{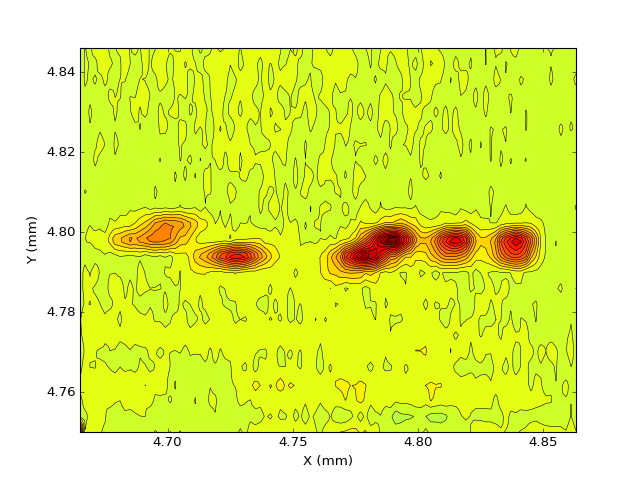}}
 \end{tabular}
\caption{\textbf{a)} The output of the 6T component imaged on the infrared camera. \textbf{b)}  A map obtained by scanning in the $XY$ direction one beam at the input of a 6-input DBC and recording the summed outputs of the components. The map is necessary for identifying the position of the inputs of the component. Once the position of the inputs is known two beams can be injected and switched  rapidly to form baselines.}
\label{JH-testbench_results}
\end{figure}

 The beam combiners were characterized on our automated test rig at Potsdam using a two-beam Michelson interferometer capable of injecting two beams through a direct and a delayed pathway. The chromatic response of the component can be characterized using two separate tunable laser sources: one at 1.3 $\upmu$m and one at 1.5 $\upmu$m. The beam from the tunable source is split using a beam splitter and directed to two kinematic mounts. The mounts can be steered using stepper motors controlled by a computer through a serial link and can address the different inputs of the component. The test bench and a map of the inputs of the components are shown in Fig.~\ref{JH-testbench} and Fig \ref{JH-testbench_results}. 

\begin{figure}[t]
\centering
\includegraphics[width=0.8\textwidth]{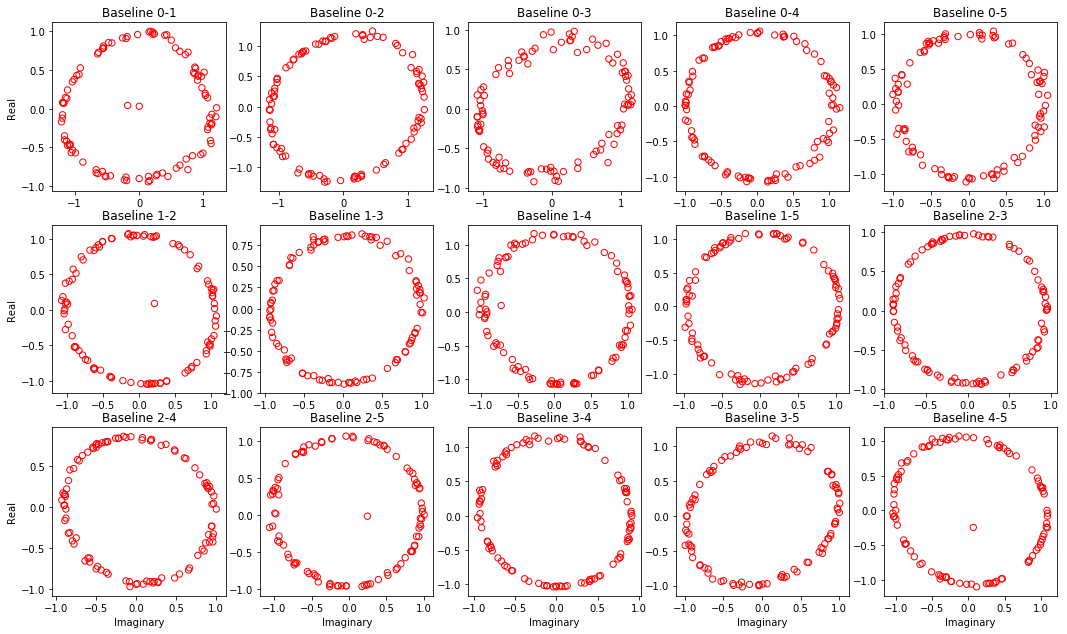} 
\caption{The  real and imaginary part of 15 visibilities obtained injecting two beams in the combiner for all possible baselines.}
\label{complVis}
\end{figure}

The inputs of the component are automatically found by performing a raster scan using the stepper motors and measuring the integrated intensity from the outputs. A $XY$ map of the intensity is then built and the outputs found.  Once we have the positions of the inputs for the two beams we can build the \ac{V2PM} matrix\cite{Saviauk2013}. One beam only is injected and all the outputs recorded by the camera. The procedure is repeated for all six beams. This gives us a measurement of the photometry of the inputs. Two beams are then injected and a USB-controlled delay line scanned with steps that are 1/10 of the laser wavelength in use. An image is acquired for each step. Sine and cosine quadratures are then fitted to the data to extract the real and imaginary part of the interfering fields. This values are placed in the \ac{V2PM} matrix together with the photometry. After the \ac{V2PM} matrix is filled up it can be inverted to give the \ac{P2VM} that relates the output recorded from the camera to the complex visibilities. Fig.~\ref{complVis} shows the real and imaginary part of 15 visibilities obtained injecting two beams in the combiner for all possible baselines.

\subsection{Exploring integrated optics beyond 3 microns}
\label{subsec:longwave}

\begin{figure}
 \begin{tabular}{ll@{}}
 \subfloat[]{\includegraphics[width=0.45\textwidth]{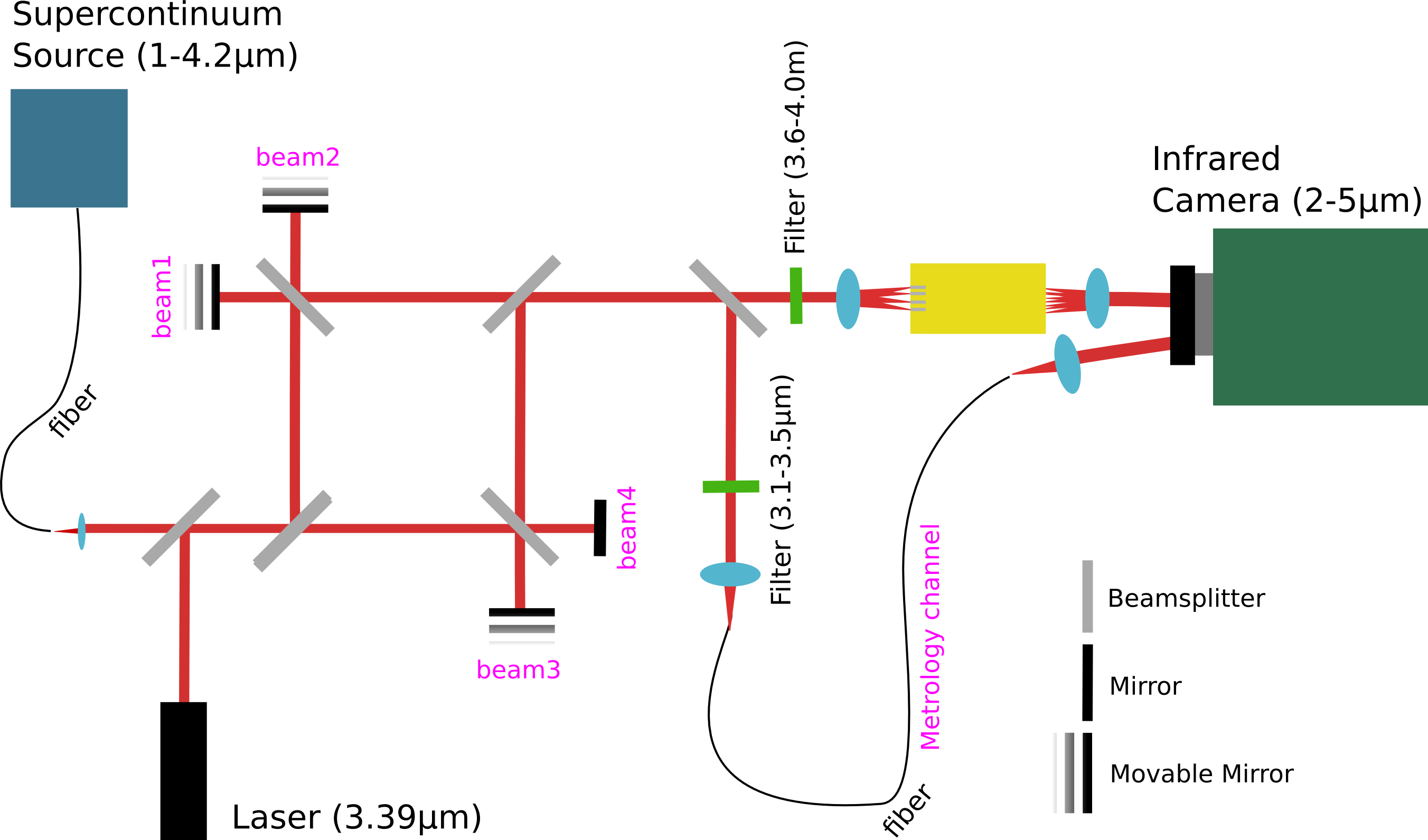} \label{fig:coherent_left} }\qquad
 & \subfloat[]{\includegraphics[width=0.45\textwidth]{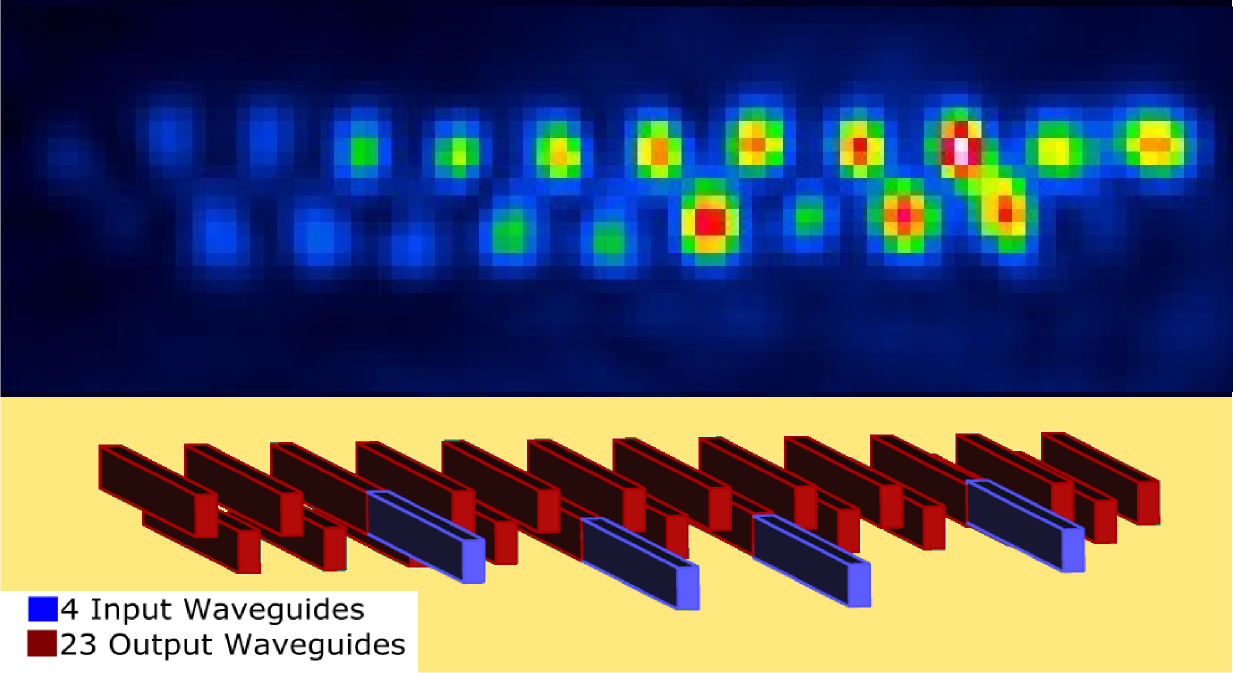} \label{fig:coherent_right}}
 \end{tabular}
\caption{\textbf{a)} 4-telescope test bench set up for the 3 to 5$\upmu$m range. The laser beam is guided through the metrology channel to calibrate the movements of the delay lines. A copy of this setup to access the 2 to 3$\upmu$m range is currently set up.
\textbf{b)} Top: Output of a 4-telescope zig-zag combiner, when all four beams are injected. The magnification is 6 and the pixel pitch of the camera is 30\,$\upmu$m. For this test, broadband light from 3.6 to 4\,$\upmu$m was used. Bottom: Schematic design of the zig-zag array.}
\end{figure}

In the context of \ac{NAIR}, the thermal infrared designates the spectral region where cryogenic instruments are required to mitigate the thermal background that would otherwise hamper any meaningful astrophysical observations. As these cryogenic instruments need to be placed in a large evacuated enclosure, they have a high cost, because of this the implementation of small scale integrated optics instruments quickly becomes an important parameter of any trade-off study. In parallel to the work conducted in the J and H bands, in \ac{NAIR} we explore solutions to cover the K (2.2 $\upmu$m) and L/M (3-5 $\upmu$m) bands leading to a unique wavelength coverage, in particular for astrophysical studies in the field of star and planet formation. To this end, we have set up a dedicated testbench to characterize components in these wavelengths ranges, see Fig.\,\ref{fig:coherent_left}.

These solutions are based on the processing of infrared glasses, transparent beyond the spectral range of the well-researched silica platform. To this end, we are experimenting with the use of \ac{ULI} \cite{Davis1996,Nolte2003,Thomson2009} in mid-IR transparent materials such as GLS and ZBLAN in order to manufacture coherent reformatter and 4-telescope beam combiners. This work is partially based on the heritage of the \ac{ALSI} project\cite{Labadie2018}, where we demonstrated the feasibility and adequate performances of laser-written integrated optics directional couplers for broadband interferometry in the L, L’ and M bands \cite{Tepper2017a,Tepper2017b}.  These devices were based on infrared GLS and ZBLAN glasses\cite{Tepper2017a,Tepper2017b} and showed a total throughput across the band of ~30 to 60\%, respectively, including Fresnel reflections and propagation losses. This result represents a critical step towards the development of the so-called ABCD 4-telescope units, which cascade several couplers in order to retrieve the interferometric visibility and phase. In one arm of the ABCD (see Ref. \cite{Benisty2009} for details) a localized increase of the refractive index allows us to insert an intrinsic $\pi$/2 phase shift in order to retrieve the instantaneous interferometric visibility and phase, without the need of scanning an internal delay-line. 2-telescope ABCD units were tested in monochromatic and polychromatic light to assess the performance of the phase-shifting unit \cite{dienerspie}.

The manufacturing of integrated optics using \ac{ULI} matches well with our research on the development of Discrete Beam Combiners targeted in \ac{NAIR}. The ability offered by \ac{ULI} to inscribe 3D structures is unique for the manufacturing of the zig-zag \ac{DBC} architecture. As we demonstrated the compatibility of \ac{ULI} with infrared glasses, the extension beyond the 1.55 $\upmu$m telecom band is quite appealing.  
Following the first characterization of \acp{DBC} at 3.39 $\upmu$m we are improving our understanding of the \ac{DBC} behavior in polychromatic light. From output images similar to those shown in Fig. \ref{fig:coherent_right}, we measure 23 interferograms (1 interferogram per output) for each of the 6 baselines (for a 4-telescope interferometer) from which an instrumental contrast and a relative phase is measured. This calibration phase allows us to derive the  \ac{P2VM} from which a static measurement of the coherence function of the source can be retrieved. The experimental characterization is complemented with a detailed simulation work using RSoft.
A key capability here for this part of the work in \ac{NAIR} is the mid-infrared test bench we have developed in Cologne, and which covers the 2--5 $\upmu$m spectral domain. It is designed to accommodate four beams that can be independently and simultaneously injected in the component. Three independent delay lines that can be scanned at different velocities allow to record temporally multiplexed signals. Finally, laser metrology operating at 3.39 microns allows us to precisely calibrate the optical delay line and to perform Fourier Transform spectroscopy of the different outputs. This will permit us to verify the level of similarity of the spectral content of each of the 23 outputs. A picture of the mid-infrared facility is shown in Fig. \ref{Koeln-testbench}.

\section{Work on incoherent reformatting devices}
\label{sec:incoherent}
 
This section details with the initial results of the incoherent side of the collaboration, where we are not trying to detect and control the phase of the system, or when we aim to scramble it completely.

The results in this section are based on 3D printed micro lens arrays, modeling of photonic devices and the development and modeling of scrambling fibers.
 
\subsection{Tip-tilt sensing using a 3D printed micro-lens array}

\begin{figure}
 \centering
 \begin{tabular}{cc@{}}
 \subfloat[]{\includegraphics[width=0.4\textwidth]{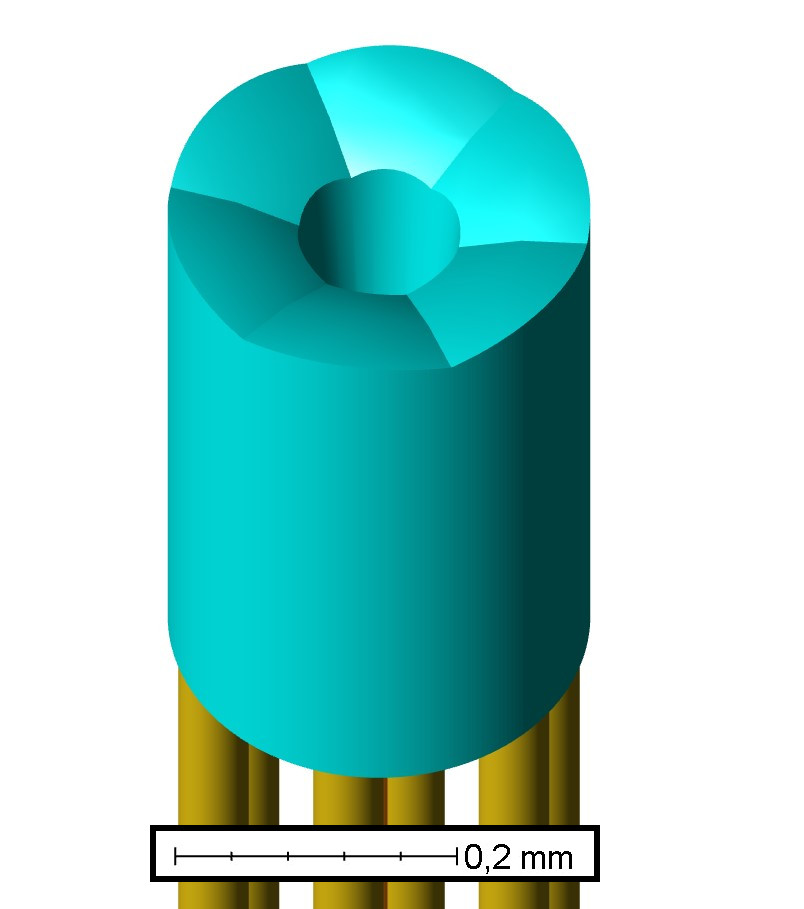} \label{fig:model}}
 & \subfloat[]{\includegraphics[width=0.4\textwidth]{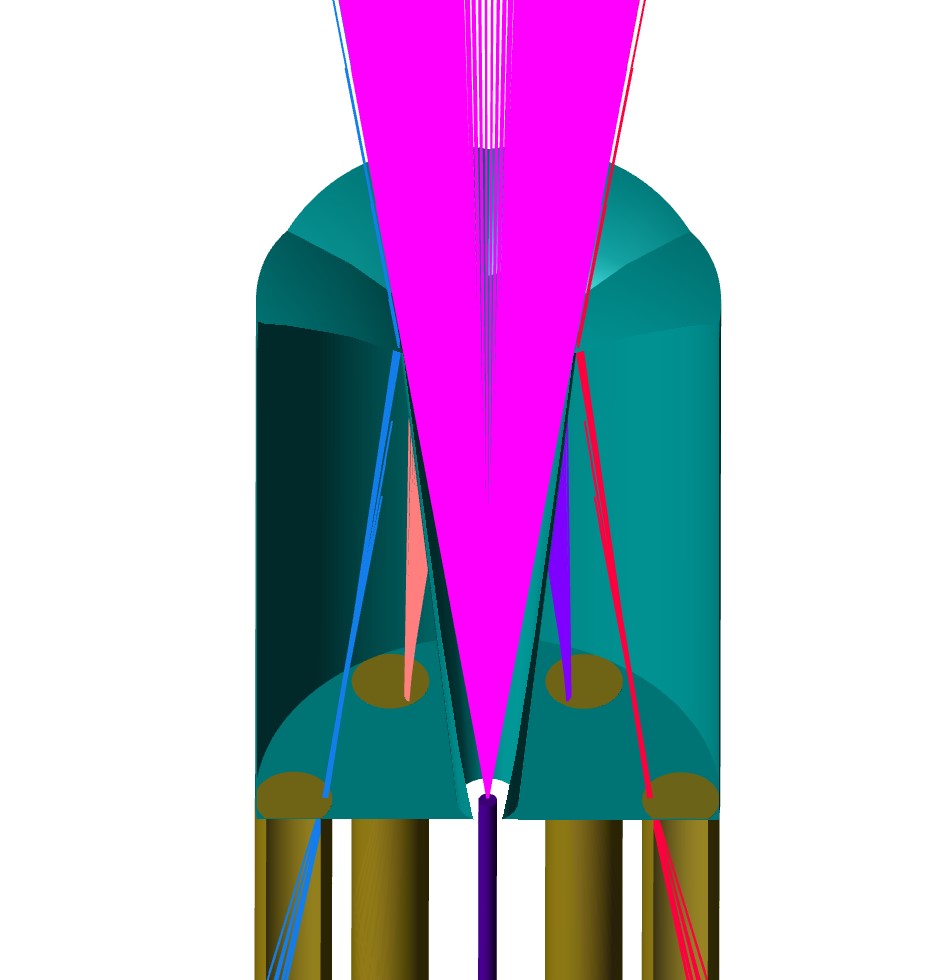} \label{fig:center}} 
 \\
 \subfloat[]{\includegraphics[width=0.4\textwidth]{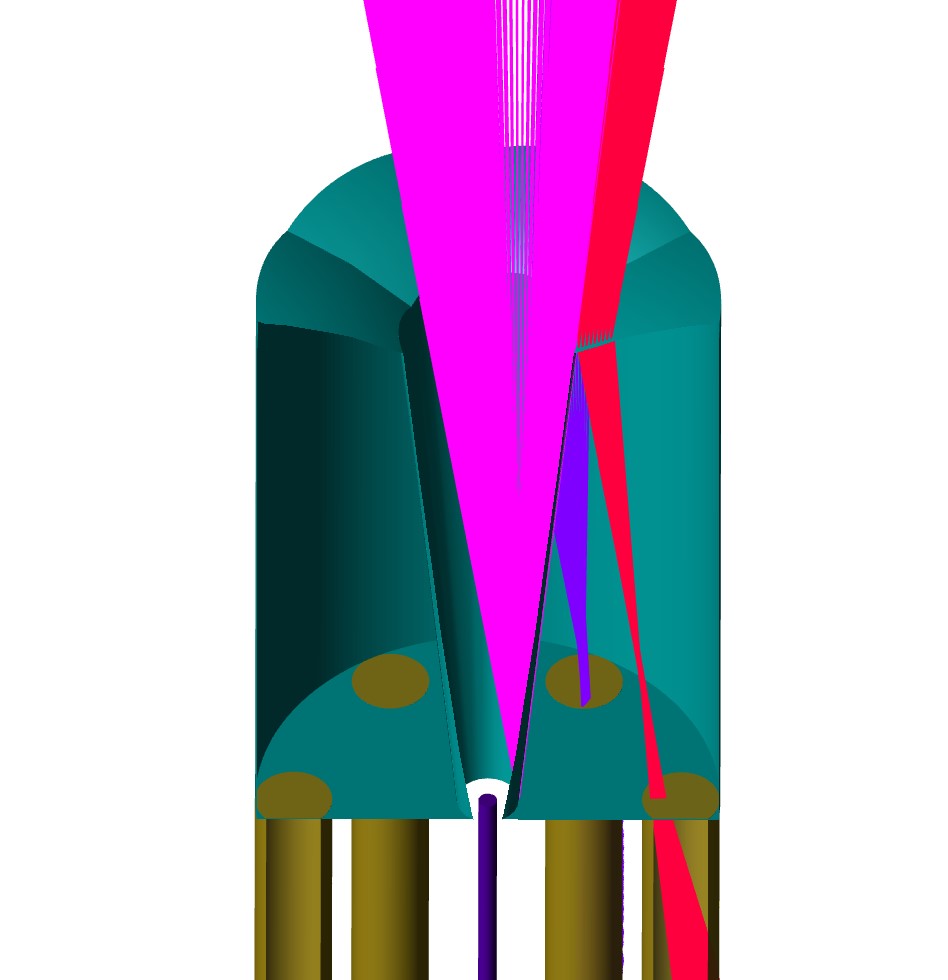} \label{fig:mis_1}}
& \subfloat[]{\includegraphics[width=0.4\textwidth]{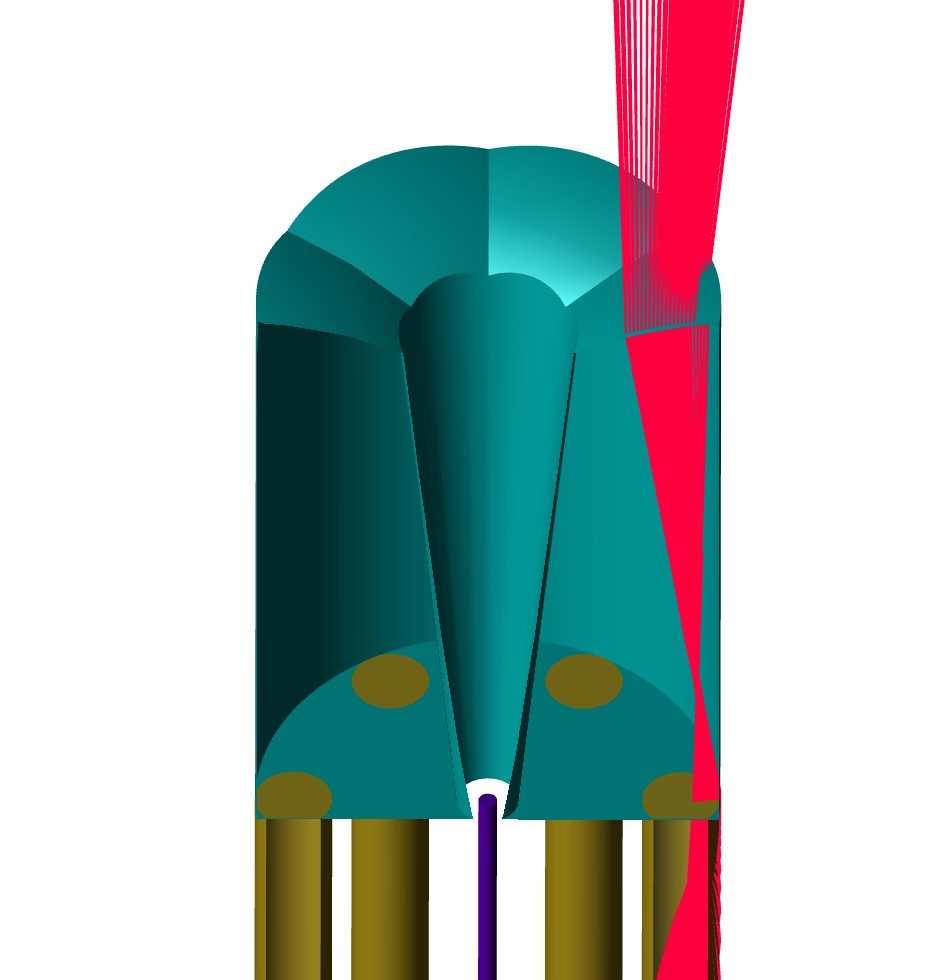} \label{fig:mis_2}}
 \end{tabular}
\caption{
3D model of the micro-lens array for tip-tilt sensor and integrated single-mode fiber coupling.
\textbf{a)} shows the surface design of the device. Cross-sections are modeled in \textbf{b)} for an aligned, \textbf{c)} for a slightly misaligned, and \textbf{d)} for a very misaligned incoming beam. While most light is coupled into the central singe-mode fiber if the beam is aligned, more light will be coupled into the outer, sensing multi-mode fiber as the beam becomes more misaligned. This signal can then be used for tip-tilt correction. For a more detailed description, see Ref. \citenum{Hottinger:2018}.
}
\end{figure}

We are also developing an integrated focal plane tip-tilt sensor for \ac{SMF} coupling.
This is creating by printing a free-form \ac{MLA} onto an array of fibers. A first study has shown the feasibility of this approach \cite{Dietrich:2017} and we have adapted it in order to fit it to the prototype iLocater fiber coupling instrument at the \ac{LBT}. Our new design consists of a central \ac{SMF} leading to the science instrument and six surrounding \acp{MMF} designed for sensing any movement of the incoming telescope beam.
The model is shown in Fig. \ref{fig:model}, with cross-sections of the \ac{MLA} with an incident aligned (Fig. \ref{fig:center}), slightly misaligned (Fig. \ref{fig:mis_1}) and badly misaligned beam (Fig. \ref{fig:mis_2}). The aligned beam is completely coupled into the \ac{SMF}, while some or most of the light is coupled into the sensing \ac{MMF} for the misaligned beams.
This design allows optimal fiber coupling if the telescope beam is aligned to the central \ac{SMF}. When the beam is decentered the light will be coupled into the surrounding sensing fibers, allowing correction.

\begin{figure}
\centering
\includegraphics[width=\textwidth]{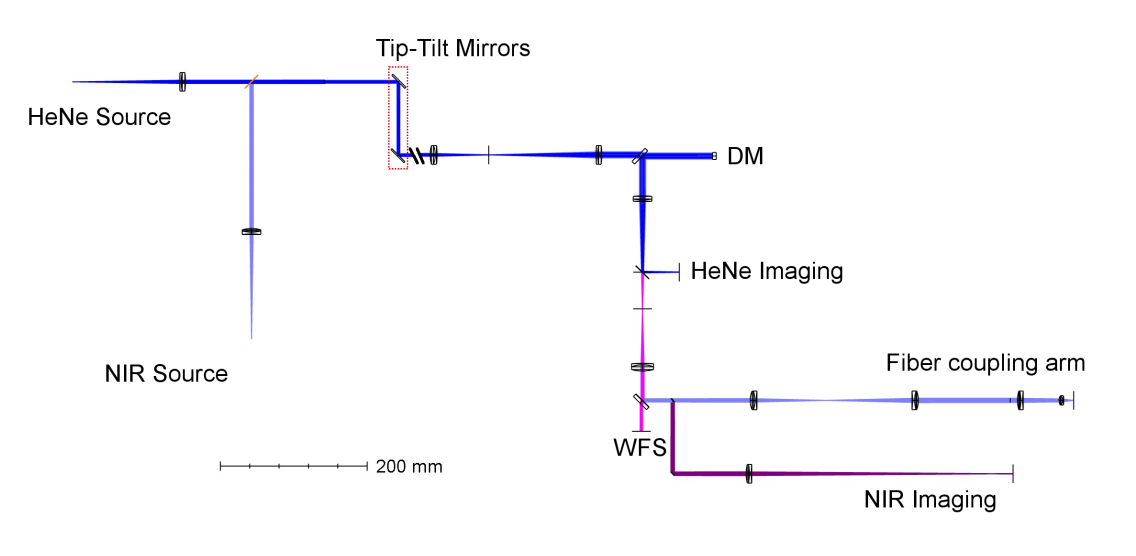} 
\caption{The design layout of the Koenigstuhl Observatory Opto-mechatronics Laboratory (KOOL). In cooperation with \ac{MPIA}, \ac{ISYS} and the \ac{LSW} we are developing a testbed for introducing and correcting vibrations and higher order aberrations. This is fed by two light sources, A HeNe laser ($\sim$633 nm) and a 1.3 $\upmu$m laser. These are combined and passed through  with two tip-tilt mirrors and a deformable mirror, allowing us to introduce and correct vibrations and aberrations. The visible light is then sent to an imaging camera and a wavefront sensor, whilst the infrared light is reflected to the imaging and fiber coupling arm where our microlens array will be placed.
For a detailed description of the setup, please see Ref. \citenum{Hottinger:2018}.}
\label{fig:KOOL_design}
\end{figure}

To prepare for on-sky tests, a testbed is also being developed in collaboration with the \ac{MPIA}, \ac{ISYS} and the \ac{LSW} is using and enhancing an \ac{AO} vibration control testbed and is called \ac{KOOL}. This testbed will allow the groups to simulate parts of the \ac{LBT}, including vibration in the system and the beam from the telescope. The optical design of the \ac{KOOL} system is shown in Fig.\ref{fig:KOOL_design}.

\subsection{Reformatter Modeling} 

\begin{figure}
 \begin{tabular}{ll@{}}
 \subfloat[]{\includegraphics[width=0.45\textwidth]{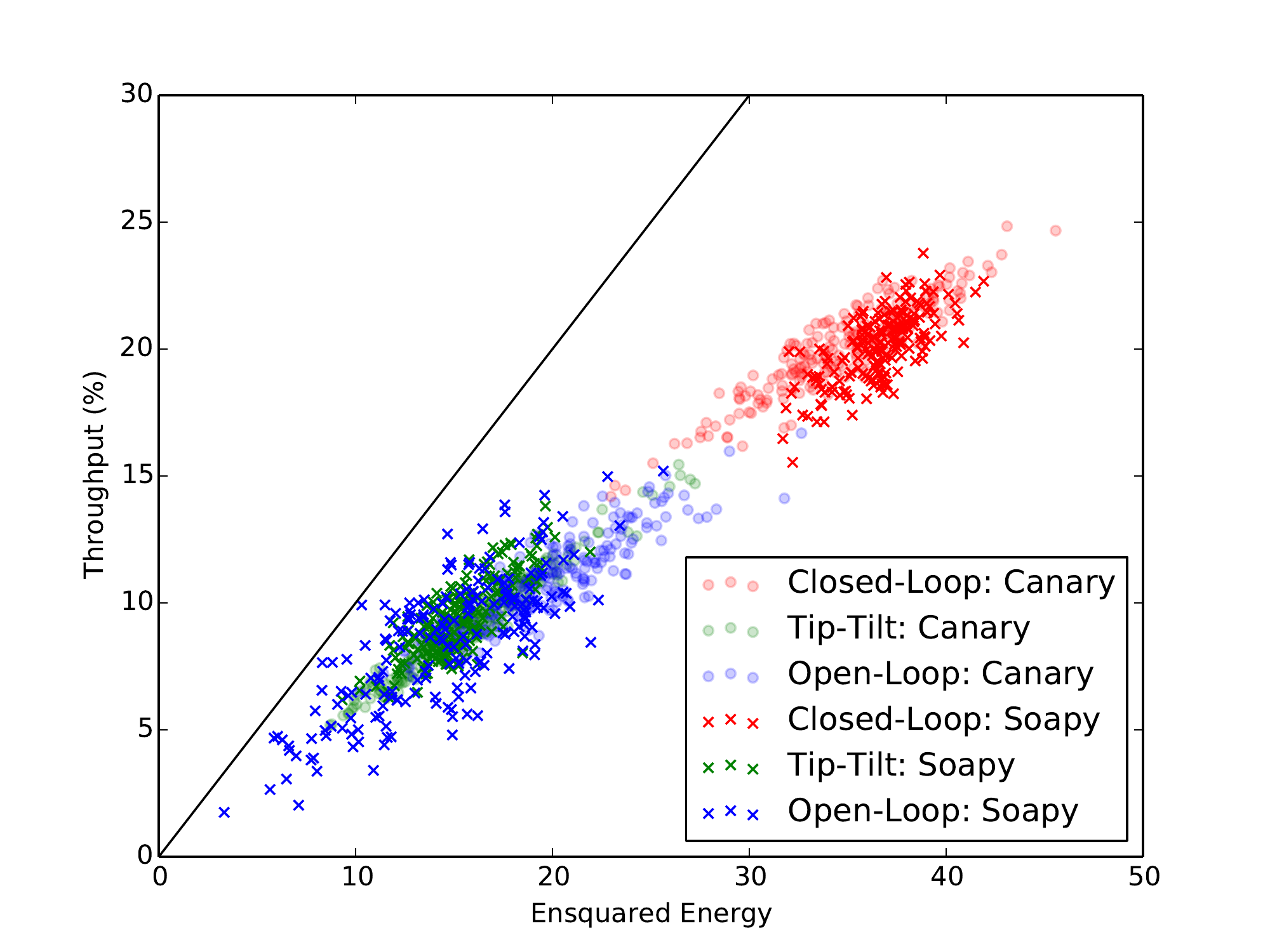} }
 & \subfloat[]{\includegraphics[width=0.45\textwidth]{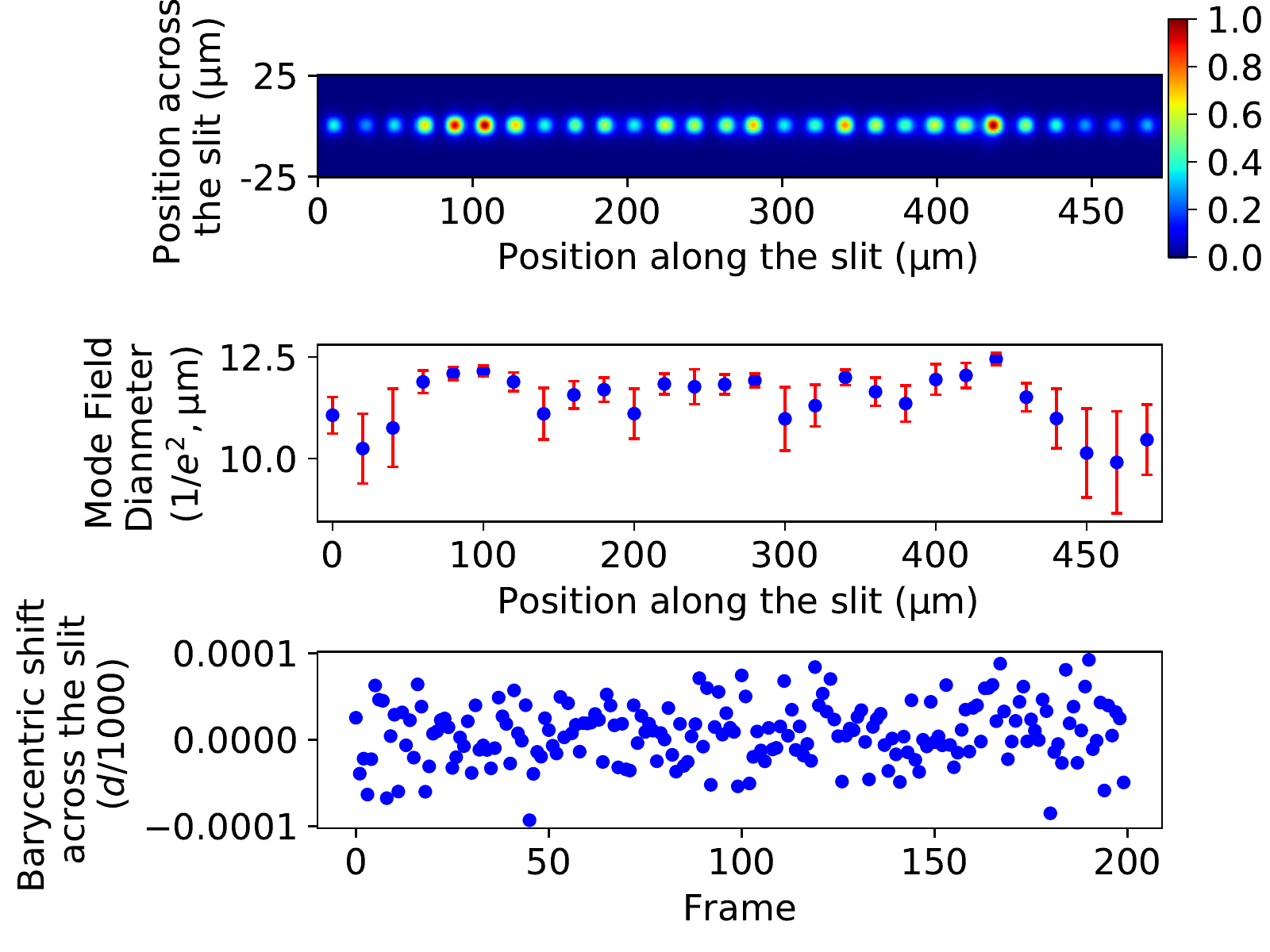}}
 \end{tabular}
\caption{\textbf{a)}: Throughput results of the
simulated photonic dicer slit end as a function of the
light portion coupled at 321 mas (the input face of the
device). Soapy (bold color) and on-sky results (transparent),
are been plotted including all three \ac{AO} operating
modes classified by color.
\textbf{b)}: Top panel: Near-field averaged (intensity)
simulations output of the optimized dicer's slit. Middle
panel: Slit's mode field diameter (MFD) shape, with 1$\upsigma$ errors derived
from individual frames. Bottom panel: Barycenter movement
across the slit derived from individual frames.}
\label{ref_prof}
\end{figure}

We have also been working on modeling photonic
reformatting devices. We have recently published a modeling
comparison with the \ac{PD} \cite{Anagnos:2018},
a device tested on-sky in 2013\cite{Harris:2015}. Our
results matched well with the reported on-sky ones.
Furthermore, we improved upon the design, improving the theoretical throughput by 6.4\% (which is a $\approx$ 30\% improvement in performance)
 and reducing the slit output movement by 50\%, resulting in lower modal noise
(see Fig. \ref{ref_prof} reproduced from Ref. \citenum{Anagnos:2018}). In addition, an improvement in the calculated value of magnification at the input of the
\ac{PD} \cite{Harris:2015}, highlighting the importance and usefulness of simulations in optimizing the throughput performance. Our study encourages detailed simulations that prove
to be a valuable tool for the design of new components for astronomy, enhancing the precision and the efficiency of them in realistic atmospheric conditions.

In this conference we will also show preliminary results from the modeling of the Hybrid reformatter \cite{MacLachlan:2017}, which compliment the reported results in literature.

\subsection{Scrambling}

Our collaboration aims to understand and improve the scrambling and modal noise properties of the \ac{PL}, in order to better stabilize their use with spectrographs. To do this we have performed the modeling detailed below.  

The traditional concept of a \ac{PL} is a device in which a series of \ac{SM} waveguides interfaces to a \ac{MM} waveguide \cite{Leon-Saval:2013}. A common method of fabricating \acp{PL} is to thermally taper a \ac{MCF}. The \ac{PL} taper transition is such that the \ac{SM} cores of the multi-core fiber become too small to guide light resulting in the cladding material containing the small residual \ac{SM} cores to form the core of the \ac{MM} section of the \ac{PL}. Although the tapered \ac{SM} cores may be too small to act as individual light guides, the remaining residual core refractive index profile does not disappear completely and therefore has to be considered when measuring the performance. The remaining residual \ac{SM} cores slightly increase the average refractive index of the \ac{MMF} core thereby increasing the \ac{NA} of the \ac{MM} \ac{PL} \cite{Birks:2015} resulting in  a possible mode number mismatch loss at the taper transition. Also the residual \ac{SM} cores appear to influence the intensity distribution across the \ac{MM} \ac{PL} core, which is an effect that needs to be fully understood in order to optimize the performance of the current \ac{MCF} \ac{PL} fiber scrambling devices \cite{Gris-Sanchez:2017}. 

\begin{figure}
\centering
	\includegraphics[width = 100 mm]{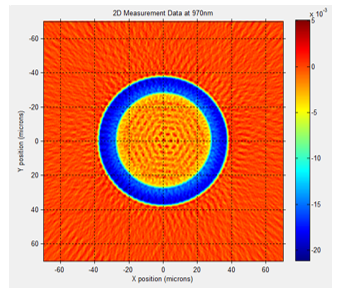}
	\caption{Cross-sectional 2D refractive index profile of a MM core of a 120 core \ac{MCF} \ac{PL}. The blue ring is the cladding of the \ac{MM} fiber section (results from fusing a fluorine doped capillary around the \ac{MCF} prior to tapering), with the yellow circle being the \ac{MM} core. The orange spots with hexagonal arrangement inside the \ac{MM} core are the residual 120 \ac{SM} cores. The scale on the right shows the refractive index variation, (relative to the refractive index of the immersion oil surrounding the fiber) measured at 970 nm.}
	\label{fig:scrambling}
\end{figure}

Fig. \ref{fig:scrambling} shows a measured 2D refractive index profile of  \ac{MM} port of a 120 core \ac{MCF} \ac{PL}. The measurement was performed using an Interfiber Analysis IFA-100 instrument. The refractive index profile of the residual \ac{SM} cores can clearly be seen as the hexagonally positioned red/orange spots within the yellow \ac{MM} core. 

In order to test the effect of residual \ac{SM} cores on light propagation through the \ac{PL}, a set of simulations were run with and without these cores. A 37 core \ac{PL} was modeled, composed of a \ac{MM} section of  length 4500 $\upmu$m and the \ac{SM} section of 8000 $\upmu$m. The \ac{MM} section included a 46.1327 $\upmu$m diameter core ($n_{\rm{core}}$ = 1.4440) surrounded by a cladding ($n_{\rm{cladding}}$ = 1.44310). The SM section contains 37 tapered cores ($n_{\rm{core}}$ =  1.45397) with a maximum diameter of 6.50 $\upmu$m at the furthest point from the \ac{MM} section. The \ac{SM} section uses the $n_{\rm{cladding}}$ = 1.4440 \ac{MM} core material as the cladding for the \ac{SM} cores, with the cores themselves increasing in size until they are large enough to guide light propagation. In the first simulation, the \ac{SM} cores taper up from 0 $\upmu$m to 6.50 $\upmu$m, and the MM core is homogeneous. In the second, 37 residual cores of 1.339 $\upmu$m diameter were included in the \ac{MM} core, with the \ac{SM} cores tapering up from this diameter.

The simulation was run by launching a Gaussian spot with a \ac{FWHM} of 8 $\upmu$m and a 1.550 $\upmu$m wavelength into the \ac{MM} end of the \ac{PL}. The spot was launched at a 15 $\upmu$m offset from the center of the fiber to better observe the transverse behavior of the propagating light.

\begin{figure} [ht]
\centering
 \begin{tabular}{cc@{}}
 \subfloat[]{\includegraphics[width=0.45\textwidth]{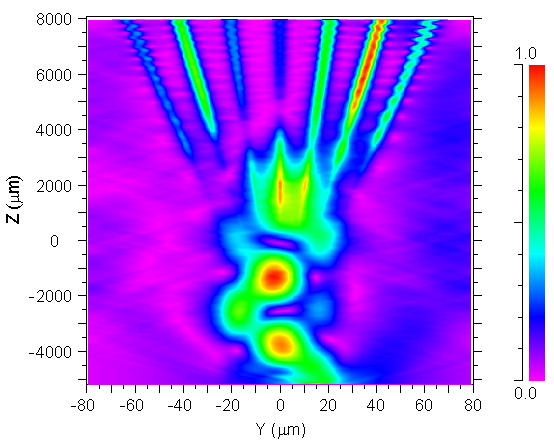} \label{fig:scrambling2a}} 
 & \subfloat[]{\includegraphics[width=0.45\textwidth]{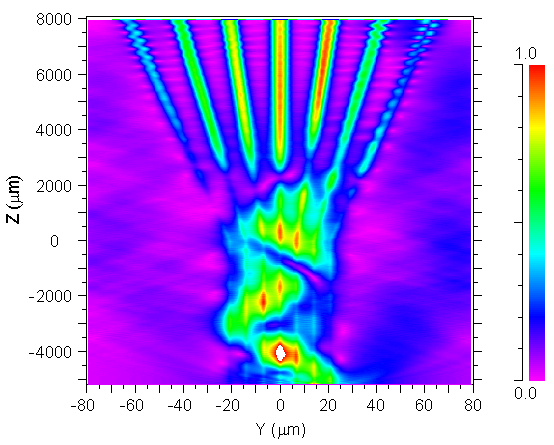} \label{fig:scrambling2b}}
 \end{tabular}
\caption{Simulation of intensity of light propagating through a 37 core \ac{PL}, shown in cross-section. A Gaussian spot with 1550 nm wavelength and 8 $\upmu$m FWHM was launched at a 15  $\upmu$m  offset. \textbf{a)} Homogeneous multi-mode section, \textbf{b)} Residual cores from fusing single-mode fibers.Color represents normalized optical intensity}
\label{fig:scrambling2}
\end{figure}

\begin{figure}[ht]
\centering
\begin{tabular}{cc@{}}
\subfloat[]{\includegraphics[width=0.45\textwidth]{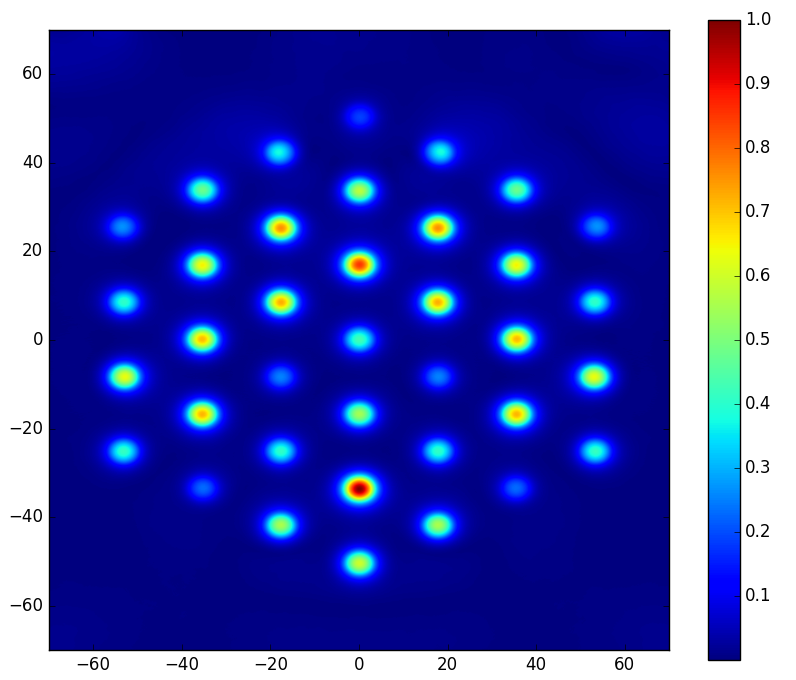} \label{fig:scrambling3a}} 
& \subfloat[]{\includegraphics[width=0.45\textwidth]{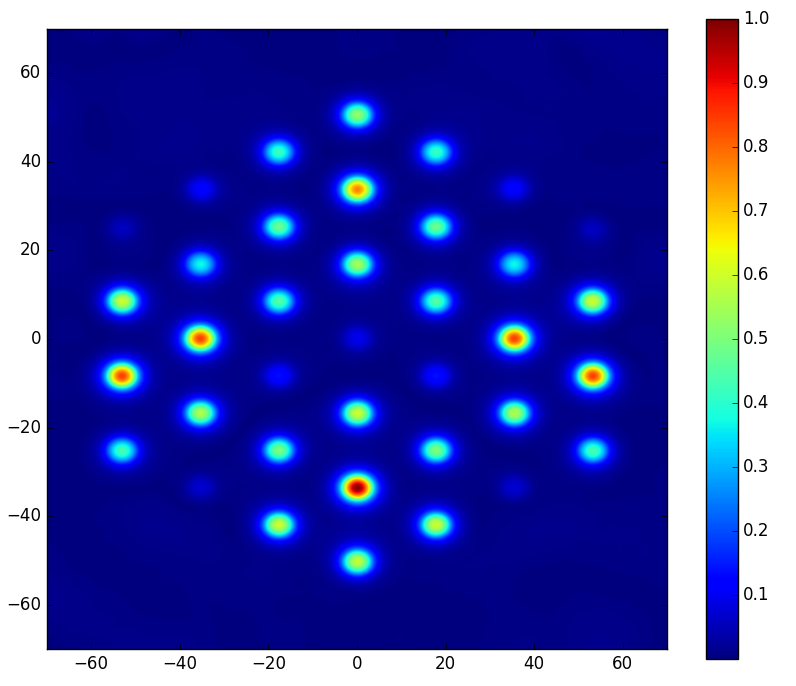}\label{fig:scrambling3b} }
\end{tabular}
\caption{Intensity output distribution at the SM end of the Photonic Lantern. \textbf{a)} Without residual cores, \textbf{b)} with residual cores. Color represents normalized optical intensity}
\label{fig:scrambling3}
\end{figure}

Fig. \ref{fig:scrambling2a} shows light propagating along the \ac{MM} section section of the \ac{PL} by total internal reflection as would be expected. When it reaches the transition region between the MM and the \ac{SM} sections, light is coupled into \ac{SM} depending on the modal distribution at that point. Figure \ref{fig:scrambling3a} shows the resulting \ac{SM} distribution, although it should be noted that this distribution is heavily dependent on factors such as fiber length and curvature which affect the modal distribution at the transition region.

Fig. \ref{fig:scrambling2b} shows the light propagation along the \ac{PL} including the residual cores. In the \ac{MM} section, light propagates with a fairly similar distribution, but an increase in intensity at the residual cores can clearly be seen and the general distribution is slightly altered. In the \ac{SM} section, the distribution is significantly altered, confirmed by Fig. \ref{fig:scrambling3b}. It can therefore be concluded that residual cores in the \ac{MM} section of a \ac{PL} caused by insufficient mixing when fusing \ac{SM} fibers cannot be ignored, and theoretical and must be considered. The small change in light propagation along the \ac{MM} section has a significant effect on the modal distribution at the transition region, which in turn significantly effects the \ac{SM} intensity distribution.

\section{Future work}
\label{sec:future}

With just over a year and a half left on the first three years of the project (there is potential for another 3 years of funding) we are on track to complete what we want. Our goals for the rest of the project are:

\begin{itemize}

\item{To fabricate the integrated 4-telescope coherent reformatter in substrate glasses such as Corning Eagle or GLS using \ac{ULI}. We will then study the deviation from the simple model that we have considered in section \ref{subsec:dbc}.}   
\item{We will also continue to develop the reformatting technologies that we successfully characterized in the lab to the telescope to perform on-sky testing. To this end, we currently investigate potential telescope sites as well the telescope-integrated optics interface. }
\item{\ac{KOOL} is currently being adapted and will soon be ready for initial tests of our new tip-tilt sensor. Once our system is tested we will utilize it with the prototype fiber feed for the high resolution spectrograph iLocater \cite{Crepp:2016}.}
\item{We have now thoroughly tested our reformatter modeling, future work will be to help optimize devices for use on sky.}
\item{The \ac{MCF} and \ac{MCF} \ac{PL} work will continue the parallel detailed development of the theory (via modeling and simulations) and experimental understanding (via laboratory testing of \ac{MCF} \acp{PL}) of the light propagation in such devices, as a function of a large range of optical input and environmental conditions. The overall aim is to improve the design and performance of \ac{MCF} based \acp{PL} with highly effective light scrambling characteristics and as integral components of reformatting devices, for applications in high precision/stability spectroscopy.}

\end{itemize}

\section{Conclusions}
\label{sec:conclusions}
\acresetall

The "\acl{NAIR}" project is a collaboration between the \acl{LSW}, Heidelberg, \acl{AIP} and the University of K\"{o}ln and is funded by the \acl{DFG}. It is developing photonic reformatting technologies for both coherent  (namely interferometry) and incoherent (namely spectroscopy) applications that will allow the next generation of astrophotonic instrumentation to function efficiently with large modern telescopes.

We have already made significant developments in modeling of the photonic lantern, reformatting devices and beam combiners. Prototypes of beam combiners, microlens arrays and photonic lanterns are also being tested.

Future work will be to continue these tests and 
integrate them into the next generation of instrumentation at observational facilities.

\acknowledgments
\acresetall

This work was supported by the \ac{DFG} through project 326946494, 'Novel Astronomical Instrumentation through photonic Reformatting'

This publication makes use of data generated at the K\"{o}nigstuhl Observatory Opto-mechatronics Laboratory (short: KOOL) which is run at the Max-Planck-Institute for Astronomy (MPIA, PI J\"{o}rg-Uwe Pott, jpott@) in Heidelberg, Germany. KOOL is a joint project of the MPIA, the Landessternwarte K\"{o}nigstuhl (LSW, Univ. Heidelberg, Co-I Philipp Hottinger), and the Institute for System Dynamics (ISYS, Univ. Stuttgart, Co-I Martin Gl\"{u}ck). KOOL is partly supported by the German Federal Ministry of Education and Research (BMBF) via individual project grants.

This research made use of Astropy, a community-developed
core Python package for Astronomy\cite{Astropy:2018},
Numpy \cite{numpy} and Matplotlib \cite{matplotlib}.

\bibliography{references} 

\begin{thebibliography}{10}

\bibitem{Hill:1980}
J.~M. {Hill}, J.~R.~P. {Angel}, J.~S. {Scott}, {\em et~al.}, ``{Multiple object
  spectroscopy - The Medusa spectrograph},'' {\em Astrophys. J. Lett.}~{\bf
  242},  L69--L72 (Dec. 1980).

\bibitem{Vanderriest:1980}
C.~Vanderriest, ``A fiber-optics dissector for spectroscopy of nebulosities
  around quasars and similar objects,'' {\em Publications of the Astronomical
  Society of the Pacific}~{\bf 92}(550),  858 (1980).

\bibitem{Leon-Saval:2005}
S.~Leon-Saval, T.~Birks, J.~Bland-Hawthorn, {\em et~al.}, ``Multimode fiber
  devices with single-mode performance,'' {\em Optics letters}~{\bf 30}(19),
  2545--2547 (2005).

\bibitem{Haniff:1987}
C.~Haniff, C.~Mackay, D.~Titterington, {\em et~al.}, ``The first images from
  optical aperture synthesis,'' {\em Nature}~{\bf 328}(6132),  694--696 (1987).

\bibitem{Perrin:2006}
G.~Perrin, S.~Lacour, J.~Woillez, {\em et~al.}, ``High dynamic range imaging by
  pupil single-mode filtering and remapping,'' {\em Monthly Notices of the
  Royal Astronomical Society}~{\bf 373}(2),  747--751 (2006).

\bibitem{Jovanovic:2012}
N.~Jovanovic, P.~G. Tuthill, B.~Norris, {\em et~al.}, ``Starlight demonstration
  of the dragonfly instrument: an integrated photonic pupil-remapping
  interferometer for high-contrast imaging,'' {\em Monthly Notices of the Royal
  Astronomical Society}~{\bf 427}(1),  806--815 (2012).

\bibitem{Haynes:2014}
D.~M. Haynes, I.~Gris-Sanchez, K.~Ehrlich, {\em et~al.}, ``New multicore low
  mode noise scrambling fiber for applications in high-resolution
  spectroscopy,'' in [{\em Advances in Optical and Mechanical Technologies for
  Telescopes and Instrumentation}{\nolinebreak\hspace{0.1em}]},   {\bf 9151},
  915155, International Society for Optics and Photonics (2014).

\bibitem{Harris:2015}
R.~J. Harris, D.~G. MacLachlan, D.~Choudhury, {\em et~al.}, ``Photonic spatial
  reformatting of stellar light for diffraction-limited spectroscopy,'' {\em
  Monthly Notices of the Royal Astronomical Society}~{\bf 450}(1),  428--434
  (2015).

\bibitem{Tuthill:2000}
P.~Tuthill, J.~Monnier, W.~Danchi, {\em et~al.}, ``Michelson interferometry
  with the keck i telescope,'' {\em Publications of the Astronomical Society of
  the Pacific}~{\bf 112}(770),  555 (2000).

\bibitem{Huby:2012}
E.~Huby, G.~Perrin, F.~Marchis, {\em et~al.}, ``First, a fibered aperture
  masking instrument-i. first on-sky test results,'' {\em Astronomy \&
  Astrophysics}~{\bf 541},  A55 (2012).

\bibitem{Minardi:2012c}
S.~Minardi, L.~Labadie, and S.~Lacour, ``Discrete optical multi-aperture
  combiner: instrumental concept,'' in [{\em SPIE Astronomical Telescopes+
  Instrumentation}{\nolinebreak\hspace{0.1em}]},   844526--844526,
  International Society for Optics and Photonics (2012).

\bibitem{Leon-Saval:2012}
S.~G. Leon-Saval, C.~H. Betters, and J.~Bland-Hawthorn, ``The photonic tiger: a
  multicore fiber-fed spectrograph,'' in [{\em Proc. of SPIE
  Vol}{\nolinebreak\hspace{0.1em}]},   {\bf 8450},  84501K--1 (2012).

\bibitem{Harris:2014}
R.~Harris, J.~Allington-Smith, D.~MacLachlan, {\em et~al.}, ``A comparison of
  concepts for a photonic spectrograph,'' in [{\em SPIE Astronomical
  Telescopes+ Instrumentation}{\nolinebreak\hspace{0.1em}]},   91474C--91474C,
  International Society for Optics and Photonics (2014).

\bibitem{Cvetojevic:2013}
N.~Cvetojevic, H.~Fernando, N.~Jovanovic, {\em et~al.}, ``High-resolution
  integrated photonic micro-spectrographs for radial velocity exoplanet
  astronomy,'' in [{\em The European Conference on Lasers and
  Electro-Optics}{\nolinebreak\hspace{0.1em}]},   CH\_1\_6, Optical Society of
  America (2013).

\bibitem{Pepe:2010}
F.~A. {Pepe}, S.~{Cristiani}, R.~{Rebolo Lopez}, {\em et~al.}, ``{ESPRESSO: the
  Echelle spectrograph for rocky exoplanets and stable spectroscopic
  observations},'' in [{\em Ground-based and Airborne Instrumentation for
  Astronomy III}{\nolinebreak\hspace{0.1em}]},  {\em SPIE Astronomical
  Telescopes+ Instrumentation} {\bf 7735},  77350F (July 2010).

\bibitem{Strassmeier:2015}
K.~Strassmeier, I.~Ilyin, A.~J{\"a}rvinen, {\em et~al.}, ``Pepsi: The
  high-resolution {\'e}chelle spectrograph and polarimeter for the large
  binocular telescope,'' {\em Astronomische Nachrichten}~{\bf 336}(4),
  324--361 (2015).

\bibitem{Quirrenbach:2010}
A.~Quirrenbach, P.~Amado, H.~Mandel, {\em et~al.}, ``Carmenes: Calar alto
  high-resolution search for m dwarfs with exo-earths with a near-infrared
  echelle spectrograph,'' in [{\em Ground-Based and Airborne Instrumentation
  for Astronomy III}{\nolinebreak\hspace{0.1em}]},   {\bf 7735},  773513,
  International Society for Optics and Photonics (2010).

\bibitem{Quirrenbach:2014}
A.~Quirrenbach, P.~Amado, J.~Caballero, {\em et~al.}, ``Carmenes instrument
  overview,'' in [{\em Ground-based and airborne instrumentation for astronomy
  V}{\nolinebreak\hspace{0.1em}]},   {\bf 9147},  91471F, International Society
  for Optics and Photonics (2014).

\bibitem{Thibault:2012}
S.~Thibault, P.~Rabou, J.-F. Donati, {\em et~al.}, ``Spirou@ cfht: spectrograph
  optical design,'' in [{\em Ground-based and Airborne Instrumentation for
  Astronomy IV}{\nolinebreak\hspace{0.1em}]},   {\bf 8446},  844630,
  International Society for Optics and Photonics (2012).

\bibitem{Zerbi:2014}
F.~Zerbi, F.~Bouchy, J.~Fynbo, {\em et~al.}, ``Hires: the high resolution
  spectrograph for the e-elt,'' in [{\em SPIE Astronomical Telescopes+
  Instrumentation}{\nolinebreak\hspace{0.1em}]},   914723--914723,
  International Society for Optics and Photonics (2014).

\bibitem{Origlia:2014}
L.~Origlia, E.~Oliva, C.~Baffa, {\em et~al.}, ``High resolution near ir
  spectroscopy with giano-tng,'' in [{\em SPIE Astronomical Telescopes+
  Instrumentation}{\nolinebreak\hspace{0.1em}]},   91471E--91471E,
  International Society for Optics and Photonics (2014).

\bibitem{Mahadevan:2014}
S.~Mahadevan, L.~W. Ramsey, R.~Terrien, {\em et~al.}, ``The habitable-zone
  planet finder: A status update on the development of a stabilized fiber-fed
  near-infrared spectrograph for the for the hobby-eberly telescope,'' in [{\em
  Ground-based and Airborne Instrumentation for Astronomy
  V}{\nolinebreak\hspace{0.1em}]},   {\bf 9147},  91471G, International Society
  for Optics and Photonics (2014).

\bibitem{GRAVITY}
{Gravity Collaboration}, R.~{Abuter}, M.~{Accardo}, {\em et~al.}, ``{First
  light for GRAVITY: Phase referencing optical interferometry for the Very
  Large Telescope Interferometer},'' {\em Astronomy \& Astrophysics}~{\bf 602},
   A94 (June 2017).

\bibitem{charles}
N.~Charles, N.~Jovanovic, S.~Gross, {\em et~al.}, ``Design of optically
  path-length-matched, three-dimensional photonic circuits comprising uniquely
  routed waveguides,'' {\em Applied optics}~{\bf 51}(27),  6489--6497 (2012).

\bibitem{Norris:2014}
B.~Norris, N.~Cvetojevic, S.~Gross, {\em et~al.}, ``High-performance 3d
  waveguide architecture for astronomical pupil-remapping interferometry,''
  {\em Optics express}~{\bf 22}(15),  18335--18353 (2014).

\bibitem{Benisty2009}
M.~{Benisty}, J.-P. {Berger}, L.~{Jocou}, {\em et~al.}, ``{An integrated optics
  beam combiner for the second generation VLTI instruments},'' {\em Astronomy
  \& Astrophysics}~{\bf 498},  601--613 (May 2009).

\bibitem{Diener2017}
R.~Diener, J.~Tepper, L.~Labadie, {\em et~al.}, ``Towards 3d-photonic,
  multi-telescope beam combiners for mid-infrared astrointerferometry,'' {\em
  Opt. Express}~{\bf 25},  19262--19274 (Aug 2017).

\bibitem{dienerspie}
R.~Diener, S.~Minardi, J.~Tepper, {\em et~al.}, ``All-in-one 4-telescope beam
  combination with a zig-zag array of waveguides,'' in [{\em Optical and
  Infrared Interferometry and Imaging V}{\nolinebreak\hspace{0.1em}]},   {\bf
  9907},  990731, International Society for Optics and Photonics (2016).

\bibitem{Arriola:2013}
A.~Arriola, S.~Gross, N.~Jovanovic, {\em et~al.}, ``Low bend loss waveguides
  enable compact, efficient 3d photonic chips,'' {\em Optics express}~{\bf
  21}(3),  2978--2986 (2013).

\bibitem{Saviauk2013}
A.~{Saviauk}, S.~{Minardi}, F.~{Dreisow}, {\em et~al.}, ``{3D-integrated optics
  component for astronomical spectro-interferometry},'' {\em Applied
  Optics}~{\bf 52},  4556 (July 2013).

\bibitem{Davis1996}
K.~M. {Davis}, K.~{Miura}, N.~{Sugimoto}, {\em et~al.}, ``{Writing waveguides
  in glass with a femtosecond laser},'' {\em Optics Letters}~{\bf 21},
  1729--1731 (Nov. 1996).

\bibitem{Nolte2003}
S.~{Nolte}, M.~{Will}, J.~{Burghoff}, {\em et~al.}, ``{Femtosecond waveguide
  writing: a new avenue to three-dimensional integrated optics},'' {\em Applied
  Physics A: Materials Science \& Processing}~{\bf 77},  109--111 (2003).

\bibitem{Thomson2009}
R.~R. {Thomson}, A.~K. {Kar}, and J.~{Allington-Smith}, ``{Ultrafast laser
  inscription: an enabling technology for astrophotonics},'' {\em Optics
  Express}~{\bf 17},  1963--1969 (Feb. 2009).

\bibitem{Labadie2018}
L.~Labadie, S.~Minardi, G.~Mart{\'i}n, {\em et~al.}, ``Progress towards
  instrument miniaturisation for mid-ir long-baseline interferometry,'' {\em
  Experimental Astronomy}  (May 2018).

\bibitem{Tepper2017a}
J.~{Tepper}, L.~{Labadie}, R.~{Diener}, {\em et~al.}, ``Integrated optics
  prototype beam combiner for long baseline interferometry in the l and m
  bands,'' {\em Astronomy \& Astrophysics}~{\bf 602},  A66 (June 2017).

\bibitem{Tepper2017b}
J.~{Tepper}, L.~{Labadie}, S.~{Gross}, {\em et~al.}, ``{Ultrafast laser
  inscription in ZBLAN integrated optics chips for mid-IR beam combination in
  astronomical interferometry},'' {\em Optics Express}~{\bf 25} (Aug. 2017).

\bibitem{Dietrich:2017}
P.-I. Dietrich, R.~J. Harris, M.~Blaicher, {\em et~al.}, ``{Printed freeform
  lens arrays on multi-core fibers for highly efficient coupling in
  astrophotonic systems},'' {\em Optics Express}~{\bf 25},  18288 (jul 2017).

\bibitem{Hottinger:2018}
P.~Hottinger, R.~J. Harris, P.-I. Dietrich, {\em et~al.}, ``Micro-lens array as
  tip-tilt sensor for single-mode fiber coupling,'' in [{\em SPIE Astronomical
  Telescopes+ Instrumentation}{\nolinebreak\hspace{0.1em}]},  International
  Society for Optics and Photonics (2018).

\bibitem{Anagnos:2018}
T.~{Anagnos}, R.~J. {Harris}, M.~K. {Corrigan}, {\em et~al.}, ``{Simulation and
  Optimization of an Astrophotonic Reformatter},'' {\em ArXiv e-prints}  (May
  2018).

\bibitem{MacLachlan:2017}
D.~G. {MacLachlan}, R.~J. {Harris}, I.~{Gris-S{\'a}nchez}, {\em et~al.},
  ``{Efficient photonic reformatting of celestial light for diffraction-limited
  spectroscopy},'' {\em \mnras}~{\bf 464},  4950--4957 (Feb. 2017).

\bibitem{Leon-Saval:2013}
S.~G. Leon-Saval, A.~Argyros, and J.~Bland-Hawthorn, ``Photonic lanterns,''
  {\em Nanophotonics}~{\bf 2}(5-6),  429--440 (2013).

\bibitem{Birks:2015}
T.~A. Birks, I.~Gris-S{\'a}nchez, S.~Yerolatsitis, {\em et~al.}, ``The photonic
  lantern,'' {\em Advances in Optics and Photonics}~{\bf 7}(2),  107--167
  (2015).

\bibitem{Gris-Sanchez:2017}
I.~Gris-S{\'a}nchez, D.~M. Haynes, K.~Ehrlich, {\em et~al.}, ``Multicore fibre
  photonic lanterns for precision radial velocity science,'' {\em Monthly
  Notices of the Royal Astronomical Society}~{\bf 475}(3),  3065--3075 (2018).

\bibitem{Crepp:2016}
J.~R. Crepp, J.~Crass, D.~King, {\em et~al.}, ``{iLocater: a
  diffraction-limited Doppler spectrometer for the Large Binocular
  Telescope},''  990819 (aug 2016).

\bibitem{Astropy:2018}
{The Astropy Collaboration}, A.~M. {Price-Whelan}, B.~M. {Sip{\H o}cz}, {\em
  et~al.}, ``{The Astropy Project: Building an inclusive, open-science project
  and status of the v2.0 core package},'' {\em ArXiv e-prints}  (Jan. 2018).

\bibitem{numpy}
S.~Van~der walt, S.~C. Colbert, and V.~Ga{\"{e}}l, ``{The NumPy array: a
  structure for efficient numerical computation},'' {\em Computing in Science
  {\&} Engineering}~{\bf 13},  22--30 (2011).

\bibitem{matplotlib}
J.~D. Hunter, ``Matplotlib: A 2d graphics environment,'' {\em Computing In
  Science \& Engineering}~{\bf 9}(3),  90--95 (2007).

\end{thebibliography}
\bibliographystyle{spiebib} 

\end{document}